\documentclass{aastex63}

\usepackage{lineno}
\usepackage[utf8]{inputenc}
\usepackage[titletoc, title]{appendix}
\usepackage{CJKutf8}
\usepackage{float}
\usepackage{blindtext}
\usepackage{gensymb}

\newcommand{\be}{\begin{equation}}
\newcommand{\ee}{\end{equation}}
\def\lta{\,\raise 0.3 ex\hbox{$ < $}\kern -0.75 em
 \lower 0.7 ex\hbox{$\sim$}\,}
\def\gta{\,\raise 0.3 ex\hbox{$ > $}\kern -0.75 em
 \lower 0.7 ex\hbox{$\sim$}\,} 
\newcommand{\xhat}{{\hat x}} 
\newcommand{\yhat}{{\hat y}} 
\newcommand{\vorb}{v_{\rm orb}}
\newcommand{\vxx}{v_{\rm x}}
\newcommand{\vzz}{v_{\rm z}}
\newcommand{\rsoi}{R_{\scriptscriptstyle SoI}} 

\def \Physics {Department of Physics, University of Michigan, Ann Arbor, MI 48109, USA}
\def \Astronomy {Department of Astronomy, University of Michigan, Ann Arbor, MI 48109, USA}
\def \Cal {Geological and Planetary Sciences, California Institute of Technology, Pasadena, CA 91125, USA}

\graphicspath{{Figures/}}

\begin{document}
\shorttitle{Capture of Interstellar Objects}
\shortauthors{Napier et al.}

\title{On the Capture of Interstellar Objects by our Solar System}

\correspondingauthor{Kevin J. Napier}
\email{kjnapier@umich.edu}

\author[0000-0003-4827-5049]{Kevin J.~Napier}
\affiliation{\Physics}
\author[0000-0002-8167-1767]{Fred C.~Adams}
\affiliation{\Physics}
\affiliation{\Astronomy}
\author[0000-0002-7094-7908]{Konstantin Batygin}
\affiliation{\Cal}

\begin{abstract}

Motivated by recent visits from interstellar comets, along with continuing discoveries of minor bodies in orbit of the Sun, this paper studies the capture of objects on initially hyperbolic orbits by our solar system. Using an ensemble of $\sim500$ million numerical experiments, this work generalizes previous treatments by calculating the capture cross section as a function of asymptotic speed. The resulting velocity-dependent cross section can then be convolved with any distribution of relative speeds to determine the capture rate for incoming bodies. This convolution is carried out for the usual Maxwellian distribution, as well as the velocity distribution expected for rocky debris ejected from planetary systems. We also construct an analytic description of the capture process that provides an explanation for the functional form of the capture cross section in both the high velocity and low velocity limits. 

\end{abstract} 


\section{Introduction} 
\label{sec:intro} 

The past few years have witnessed the detection of two interstellar bodies passing through the solar system on hyperbolic orbits. The discoveries of the irregular body ‘Oumuamua \citep{meech2017} and the comet Borisov \citep{jewitt2019} sparked immediate interest in characterization of these objects and facilitated wide-ranging speculation regarding the possibility that our solar system is more broadly contaminated by minor bodies of extra-solar origin (e.g.,  \citealt{siraj2019,namouni}). Although no current evidence indicates that any specific objects in the solar system are of extrinsic origin \citep{morby2020}, the question of whether or not any such objects reside in interplanetary or trans-Neptunian space is of considerable interest. Motivated by these issues, this paper reconsiders the capture of external bodies by our solar system. The calculation of the capture cross sections is the first step in assessing whether or not the solar system presently contains  quasi-permanently trapped interstellar bodies. This treatment also provides constraints on the expected orbits of any such material.

The dynamics of the outer solar system represents one of the oldest problems in theoretical astrophysics. Starting more than two centuries ago, classic studies include the long-term stability of the solar system \citep{lagrange1776,laplace1799}, the origin of comets \citep{laplace1806}, and orbital anomalies that led to the discovery of Neptune \citep{leverrier,adams1846}. Over recent decades, the outer solar system has  revealed itself to be increasingly complicated, with the discovery of the Kuiper Belt \citep{luujewitt}, dozens of dwarf planets (starting with \citealt{sedna}, or perhaps \citealt{tombaugh}), high-inclination objects \citep{bp519}, and aligned extreme trans-Neptunian objects \citep{sheptruj2016} that led to the hypothesis of a possible ninth planet \citep{batbrown,pninereview}. The more recent discovery of interstellar objects  \citep{meech2017,jewitt2019} adds to the intrigue.  Both the complex orbital architecture of the solar system and the presence of interloping objects motivates this present study. The goal is to determine cross sections for the capture of foreign bodies by the solar system, and to obtain a deeper understanding of the capture process. 

The possible capture of interstellar bodies by the solar system also has a long history. The general problem of interacting binaries was considered by \cite{heggie1975}, where the subset of `resonant' encounters lead to capture. Subsequent studies have carried out numerical explorations of the capture process  specifically for our solar system, often considering only the Sun-Jupiter system (see, e.g., \citealt{valtonen1982,valtonen1983,siraj2019}). Additional studies consider capture for specific scenarios, including capture by compact objects \citep{pineault}, capture of interstellar objects from the field \citep{lingamloeb,hands2020}, the formation of wide binaries \citep{kouwenhoven}, and the possible capture of Planet Nine  \citep{liadams2016,mustill}. Most of these previous studies calculate the capture rate by sampling a given distribution of encounter speeds between the incoming body and the solar system. These studies generally use the field star velocity distribution, with dispersion $\sim40$ km/s \citep{binneytremaine} or that appropriate for the solar birth cluster \citep{zwart2009,adams2010,pfalzner2013,parker2020}, where the velocity dispersion is expected to be $\sim1$ km/s \citep{ladalada}. Notice, however, that the velocity distribution for rocks (or planets) ejected by solar systems will not generally have a simple Maxwellian form.\footnote{As one example, the distribution of speeds for planets ejected from crowded solar systems has the approximate form $dP/dv=4v/(1+v^2)^3$ (e.g., \citealt{moorhead}).}  

The objective of this paper is to extend the aforementioned previous work concerning the capture of interstellar bodies by the solar system. Whereas most studies determine capture rates and cross sections for a given distribution of velocities, this work finds the cross section $\sigma(v_\infty)$ as a function of relative velocity. The results can then be integrated (after the fact) for any distribution of velocities of interest. This approach is much more computationally expensive than previous treatments, but is made possible with current computational capabilities. Specifically, this paper reports the results from $\sim5\times10^8$ fly-by simulations. In addition, we carry out the simulations for solar systems models including all four giant planets. Although earlier work \citep{heggie1975,pineault} provides analytic estimates for the cross sections, exact forms are not available (primarily due to the lack of an analytic solution to the gravitational three-body problem). We revisit this issue using a different (but equivalent) set of approximations. We then compare the numerical and analytic results for the cross section as a function of velocity, and find good agreement. 

\section{Dynamics of the Rock Capture Process} 
\label{sec:capture} 

This section presents an analytic description of the rock capture process. The capture of an incoming body occurs through the time dependence of the gravitational potential of the solar system.  In this treatment, we consider the incoming orbit in two regimes. In the outer regime, at large distances, the rock executes a hyperbolic orbit about the center of mass of the solar system. In the inner regime, at closer distances, the rock can enter into the sphere of influence of individual solar system members (e.g., the Sun or Jupiter), and then be described by a hyperbolic orbit around that body. Under favorable conditions, the deflection by the solar system body during the close encounter can lead to energy loss and capture in the center of mass frame. This effect is essentially the inverse of the gravitational slingshot mechanism by which satellites are boosted through planetary encounters. Note that by dividing the orbit into two regimes, we are implicitly assuming that 3-body effects are not important.

For the sake of definiteness, we consider only one planet at a time, and work in the limit where the masses of the rock $\mu$, the planet $m$, and the star $M$ obey the ordering 
\be
\mu \ll m \ll M\,.
\label{ordering} 
\ee
The incoming orbit of the rock is characterized by its asymptotic speed $v_\infty$ and impact parameter $b$. For given input variables $(v_\infty,b)$, we can define the orbital elements and related physical quantities, including the specific energy and angular momentum, 
\be
E = {1\over2} v_\infty^2 \qquad {\rm and} \qquad 
J = b v_\infty \,,
\ee
the semi-major axis and eccentricity, 
\be
|a| = -a = {GM \over v_\infty^2} \qquad {\rm and} \qquad 
e^2 = 1 + b^2/a^2 \,,
\ee
and the perihelion distance 
\be
r_p = p = a(1-e) = |a| (e-1)  \,.
\label{perihelion} 
\ee
Note that, by convention, the semi-major axis $a<0$. To fully characterize the orbit, one must also specify the inclination angle of the incoming trajectory.

It is useful to define the effective cross section for hyperbolic orbits to enter the giant planet region of the solar system. In order for the incoming rock to experience the time-dependence of the gravitational potential, the perihelion $r_p$ must be smaller than the semi-major axis $a_p$ of the planet of interest. This condition implies that the impact parameter $b$ is bounded from above by $b^2 \le a_p^2 + 2 a_p |a|$, where $a$ is the semi-major axis of the incoming orbit. The nominal cross section $\sigma_0$ for orbit crossing is thus given
by
\be
\sigma_0 = \pi \left[ a_p^2 + 2 a_p |a| \right] \approx 
2\pi {GM \over v_\infty^2} a_p \,, 
\label{sigmazero} 
\ee
where the final equality holds for essentially all incoming speeds of interest ($v_\infty^2<GM/a_p$). The capture cross section will be some fraction of the fiducial cross section (\ref{sigmazero}).\footnote{Note that the interpretation of this fiducial cross section would be more complicated if the planetary orbit had significant eccentricity. Nonetheless, one can always scale the results to the expression of equation (\ref{sigmazero}).} 

\subsection{Gravitational Slingshot Mechanism for Close Encounters} 
\label{sec:slingshot} 

\begin{figure}[h]
\centering
\includegraphics[width=1\textwidth]{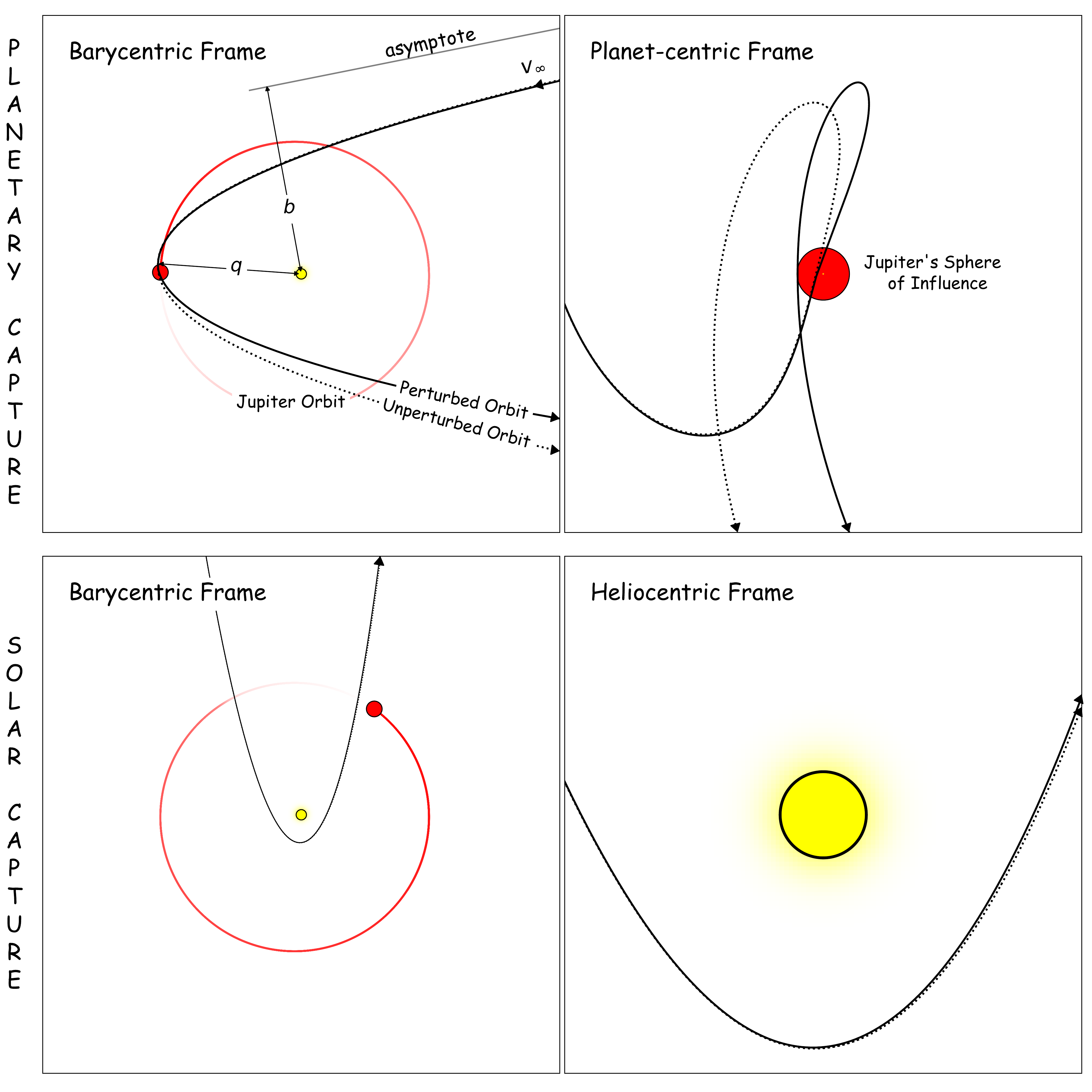}
\caption{Examples of capture events. In each frame, the dotted black line denotes the rock's initial (unbound) orbit; the solid black line denotes the rock's trajectory after a perturbation; and the arrows specify the direction of the orbit. The yellow circle represents the Sun, the filled red circle represents Jupiter's sphere of influence drawn at the epoch of the rock's closest approach, and the red line marks Jupiter's orbit. The top row shows a capture by Jupiter: the left panel is in the frame of the solar system's barycenter, and the right panel is in Jupiter's rest frame. Note that the rock gets well inside of Jupiter's sphere of influence, but does not actually collide with the planet. The bottom row shows a capture by the Sun: the left panel is in the frame of the solar system's barycenter, and the right panel is in the Sun's rest frame. If the target (Sun or planet) has a component $U$ of its velocity moving away from the incoming rock as it approaches periapsis, then the encounter causes the rock to lose energy in the inertial reference frame, thereby allowing the rock to potentially enter into a bound orbit.}  
\label{fig:encounter} 
\end{figure}

For the inner regime defined above, we consider close encounters of incoming rocks on initially hyperbolic trajectories with much larger target bodies (either the Sun or one of the giant planets). We can define the coordinate system so that the rock approaches the target body from the $+\xhat$ direction and from the $+\yhat$ direction, where the angle of the incoming trajectory is $\theta$ in the center of mass frame (see the first panel in Figure \ref{fig:encounter}). The rock initially has speed $v_\infty$ in the inertial reference frame and the target body has speed $U$. In the frame of the target, the incoming rock has velocity
\be
{\bf v}_{1tar} = \left( - [v_{x} - U] , v_y \right) 
\ee
and the outgoing velocity has the form 
\be
{\bf v}_{2tar} = \left( + [v_{x} - U] , v_y \right) \,.
\ee
This second equation assumes that the encounter is symmetric, i.e., the outgoing trajectory of the rock is the mirror image of the incoming trajectory. This approximation thus assumes that the larger body does not change its velocity (consistent with the ordering of  equation [\ref{ordering}]) and that the encounter time is short compared to the orbital period.  In the center of mass reference frame, the incoming velocity has the form 
\be
{\bf v}_{1cm} = \left( - v_{x}, v_y \right) 
= \left( - v \cos\theta, v \sin\theta \right) \,,
\ee
whereas the outgoing velocity becomes 
\be
{\bf v}_{2cm} = \left( + [\vxx - 2U], v_y \right) 
= \left( + [v \cos\theta - 2U], v \sin\theta \right) \,.
\ee
The final speed is then given by the expression 
\be
v_2^2 = [v \cos\theta - 2U]^2 + v^2 \sin^2\theta 
= v^2 - 4Uv \cos\theta + 4U^2 \,. 
\label{vtwocm} 
\ee
Note that this discussion assumes that the encounter is symmetric in the reference frame of the capturing body. This approximation is expected to be valid because only close encounters with the body result in capture events, and such close encounters will be symmetric to leading order.

The discussion thus far has implicitly assumed that the target is moving in the $-\xhat$ direction. In general, however, the target can also have a $\yhat$ component to its velocity. Because of the geometry of the encounter, however, only the $\xhat$ component of the rock velocity changes (in this approximation). We can thus incorporate the more general case by interpreting the velocity $U$ as the component of the target velocity in the $-\xhat$ direction. With this definition of  $U$, the final speed still obeys equation (\ref{vtwocm}). 

\subsection{Solar Close Encounters} 
\label{sec:starslingshot} 

When the rock is far away from the Sun,  it effectively orbits the center of mass of the system. When the radial distance of the rock becomes sufficiently small, however, its orbit is determined by the location of the Sun. We can delineate the boundary between these two regimes by considering the acceleration (and hence forces) in an accelerating reference frame. As expected, the orbit of Jupiter provides an approximate boundary between the outer problem (hyperbolic orbit about the center of mass of the solar system) and the inner problem (close encounter with the Sun).

When the rock enters the sphere of influence of the Sun, its speed is given by 
\be
v^2 = v_\infty^2 + {2GM \over r} \,,
\label{v2sun} 
\ee
where $r<a_J$ is the location of the rock. It will then execute a (hyperbolic) orbit around the Sun. Due to the motion of the Sun about the center of mass of the solar system, the post-encounter velocity will be given by 
\be
v_2^2 = v^2 - 4Uv \cos\theta + 4U^2 \,, 
\ee
where $U$ is the component of the solar velocity in the direction of the perihelion of the orbit and $\theta$ defines the shape of  the hyperbola. Capture of the rock requires that  $v_2^2<v^2-v_\infty^2$, so that we obtain the constraint 
\be
v_\infty^2 < 4Uv \cos\theta - 4U^2 \,. 
\label{vlimit} 
\ee
Here, the angle $\theta$ is determined by the parameters of the original hyperbolic orbit about the Sun, so that 
\be
\cos\theta = {1 \over e} = {|a| \over (a^2 + b^2)^{1/2}} \,,
\label{deftheta} 
\ee
where $a$ is the semi-major axis and $b$ is the impact parameter. The capture constraint thus becomes 
\be
v_\infty^2 < 4 \alpha {m \over M} \left({GM\over a_J}\right)^{1/2} 
\left[ v_\infty^2 + {2GM \over r_p} \right]^{1/2} 
{|a| \over (a^2 + b^2)^{1/2}} + 
{\cal O} \left( {m^2\over M^2} \right) \,, 
\ee
where we have assumed that the speed $U$ is some fraction of the speed of the Sun in its orbit about the center of mass. Specifically, we define the parameter $\alpha$ such that 
\be
U \equiv \alpha {m \over M} \left({GM\over a_J}\right)^{1/2} \,,
\ee
where $m$ is the mass of Jupiter.  Since the speed of the incoming rock $v\gg{U}$ for the close encounters of interest, we ignore the $U^2$ term. Finally, we evaluate the rock velocity at the perihelion position $r_p$ (see equation [\ref{perihelion}]), as this location
corresponds to where the close encounter takes place.  Working to consistent order, the expression for the capture constraint can be written in the form 
\be
v_\infty^2 < 4 \alpha {m \over M} \left({GM\over a_J}\right)^{1/2} 
\left[ {4GM|a| \over b^2} \right]^{1/2} = 
8 \alpha {m \over M} \left({GM\over a_J}\right)^{1/2} 
{GM \over b v_\infty} \,.
\label{vinflimit} 
\ee
The constraint can be written as a limit on the impact 
parameter, i.e., 
\be
b < 8 \alpha {m \over M} \left({GM\over a_J}\right)^{3/2} 
\, v_\infty^{-3} a_J \approx 93\,\,{\rm au}\,\,\alpha 
\left({v_\infty\over1\,\,{\rm km/s}}\right)^{-3}\,.
\label{bmax} 
\ee
If one requires that the rocky body is not only captured, but is captured into an orbit with semi-major axis less than some maximum value $a_{\rm max}$, then the left-hand-side of equation (\ref{vinflimit}) can be replaced with $v_\infty^2 + \vxx^2$,  where $\vxx^2\equiv$ $GM/a_{\rm max}$. Finally, note that this treatment implicitly assumes that $U>0$. If the Sun is moving in the opposite direction, the encounter would cause the incoming rocky body to gain energy, and capture does not take place.

Given the approximations presented above, the resulting cross section for capture can be written in the from 
\be
\sigma = 64 \pi \alpha^2 a_J^2 \left({m \over M}\right)^2
\left({GM\over a_J}\right)^3 
{1 \over v_\infty^2 (v_\infty^2 + \vxx^2)^2}  \,. 
\label{sunsigma} 
\ee
This cross section is specified up to the constant $\alpha$, which is expected to be of order (but less than) unity.  This form is consistent with those derived earlier by other means \citep{heggie1975,pineault,valtonen1983}. Notice that this derivation breaks down for sufficiently high incoming speeds, $v_\infty\gta8$ km/s, as shown in Appendix \ref{sec:bound}.

\subsection{Planetary Close Encounters} 
\label{sec:planslingshot} 

Another channel for capture occurs through close encounters with the giant planets, most often Jupiter, which will be considered in this discussion.  Equation (\ref{sigmazero}) represents the cross section for an incoming rock to enter the sphere of radius $a_J$. Only a fraction of the incoming trajectories $f_1 = \rsoi^2/4a_J^2$ will enter the sphere of influence of Jupiter,\footnote{Note that the sphere of influence, as defined here, corresponds to the location where the incoming trajectory switches from a two-body problem with
central mass $M$ to a two-body problem with central mass $m$ in the matched conics approximation. The boundary $\rsoi$ is comparable to, but not equivalent to, the Hill radius $R_H=a(m/3M)^{1/3}$.}
delineated by $\rsoi\approx{a_J}(m/M)^{2/5}$ \citep{batebook}.  
However, not all of the orbits that enter the sphere of influence will pass close enough to the planet to experience significant deflection. As a result, we must estimate the smaller fraction $f_2$ of trajectories that allow for capture.

As a rough approximation, significant deflection requires $\cos\theta$ to be of order (but still less than) unity, which in turn implies $b_{hp}\sim|a|_{hp}$ (equation [\ref{deftheta}]), where $(a_{hp},b_{hp})$ correspond to the elements
of the hyperbolic orbit around the planet.  When the rock encounters the planet, its speed in the solar reference frame is given by equation (\ref{v2sun}) evaluated at $r\approx a_J$. The asymptotic speed $(v_\infty)_{hp}$ for the hyperbolic orbit about the planet depends on the planetary motion, but will typically be of the same order.  We can thus write 
\be
(v_\infty^2)_{hp} = v_\infty^2 + \beta {GM \over a_J} \equiv 
v_\infty^2 + \vzz^2 \,,
\ee
where $\beta$ is a dimensionless factor of order unity and where the second equality defines the velocity scale $\vzz$. The semi-major axis of the hyperbolic planetary encounter is given by 
\be
(|a|)_{hp} = {G m \over (v_\infty^2)_{hp} } = {Gm \over 
v_\infty^2 + \vzz^2} \sim {m\over M} a_J \,. 
\ee
Since we require $b_{hp}\lta|a|_{hp}$ and $|a|_{hp}\ll\rsoi$, the fraction $f_2=a_{hp}^2/4 a_J^2$. The resulting cross section for capture due to planetary encounters has the form 
\be
\sigma = {\gamma \pi \over 2} {GM \over v_\infty^2} a_J 
\left( {Gm \over a_J \left( v_\infty^2 + \vzz^2 \right)} \right)^2 
= {\gamma \pi a_J^2 \over 2} \left({m\over M}\right)^2 
\left( {GM \over a_J} \right)^3
{1 \over v_\infty^2 \left( v_\infty^2 + \vzz^2 \right)^2} \,, 
\label{plansigma} 
\ee
where we have introduced a dimensionless factor $\gamma$ that is expected to be of order unity.  Note that this expression has a form similar to that of equation (\ref{sunsigma}), which corresponds to the capture cross section for solar encounters. Keep in mind, however, that the velocity scales are different and are expected to obey the ordering $\vxx<\vzz$.

\subsection{Energy Distribution of Newly Bound Orbits} 
\label{sec:energy} 

Using the results from the previous section, we can write the  post-encounter speed of the rock in the form 
\be
v_2^2 \approx v^2 - 4Uv \cos\theta\,.
\ee
The semi-major axis $a_b$ of the bound orbit is defined so that 
\be
{GM \over a_b} = 2 {GM\over r} - v_2^2 = 
4Uv \cos\theta - v_\infty^2 \,. 
\ee
Let us now define a scale length $b_0$ according to 
\be
b_0 \equiv 8 {m \over M} \left({GM\over a_J}\right)^{3/2} 
{a_J \over v_\infty^3} \alpha\, \sim 100\,\, {\rm au} 
\left({v_\infty\over1\,{\rm km/s}}\right)^{-3} \,. 
\ee
With this construction, the semi-major axis of the bound orbit is given by 
\be
{GM \over a_b v_\infty^2} = {b_0 \over b} - 1 
\qquad \Rightarrow \qquad a_b = {a_0 b \over b_0-b} = 
{a_0 \chi \over 1 - \chi} \,, 
\label{atochi} 
\ee
where we let $a_0=|a|$ denote the (magnitude of) the semi-major axis of the initial hyperbolic orbit, and where the final equality defines $\chi\equiv{b}/b_0$. The criterion for obtaining a bound orbit (from the previous section) is equivalent to the requirement $b<b_0$ $(\chi<1)$. Since the cross section  depends on $b^2$, the distribution of impact parameters will be weighted towards larger values.  This finding, in turn, implies that typical bound orbits will have final semi-major axes comparable to the starting (negative, hyperbolic) semi-major axis of the incoming orbit. For $v_\infty=1$ km/s, for example, bound orbits will typically have $a_b\sim1000$ au. In order to obtain tighter orbits comparable to the size of the solar system  (or even the Kuiper belt), we need $a_b\lta100$ au, which in turn implies that $b\lta b_0/10\sim10$ au. 

If we assume that the impact parameters $b$ are uniformly distributed over an area, with a maximum value $b_0$, then the probability distribution for the dimensionless quantity $\chi$ has the simple form $dP = 2 \chi d\chi$. Using equation (\ref{atochi}), we can determine the probability distribution for the semi-major axes of the bound orbits, i.e.,
\be
{dP \over da_b} = {2 a_b a_0 \over (a_b + a_0)^3} \,. 
\label{adistrib} 
\ee
As written, this distribution is normalized over the interval $0<a_b<\infty$.

Note that the distribution of equation  (\ref{adistrib}) corresponds to the semi-major axes of the bodies when they are captured. The orbital elements of the captured objects will continue to evolve (e.g., through continued close encounters with the planets), so that quasi-stable orbits will display a different distribution (which should be explored further in future work).

\section{Numerical Results}
\label{sec:simulaitons}

The cross sections derived in the previous section made use of a number of approximations. In this section we use a suite of more than 500 million simulations to numerically compute the capture cross section.

\subsection{Simulation Details}

We sample rocks of mass $10^{-9}$ $M_{\odot}$ isotropically on the sphere at a barycentric distance of $10^{9}$ au. Each rock's velocity unit vector is uniquely defined by its position on the sphere, pointing directly toward the solar system barycenter. We then assign each rock an impact parameter at some random angle in its plane tangent to the sphere. We randomly sample the impact parameter uniformly given the condition that the maximum pericenter distance $q_{\text{max}} \leq 12$ au---comfortably above the largest pericenter distance for capture not attributable to chance close encounters with a giant planet. Finally we scale the rock's velocity unit vector by a factor
\begin{equation}
    v = \sqrt{v_{\infty}^2 + \frac{2\mu}{r}}
    \label{eq:v_correction}
\end{equation}
where $\mu = G\sum_i m_i$ and $i \in \{ \text{Sun, Jupiter, Saturn, Uranus, Neptune} \}$. In Equation \ref{eq:v_correction}, $v_{\infty}$ is the rock's field (or cluster) velocity at infinity, and the second term accounts for the kinetic energy that the rock gains by falling from infinity to a barycentric distance $r$. 

The above procedure gives us a state vector, from which we compute a body's Keplerian orbital elements. To save computation time, we use these elements to propagate each rock along its unperturbed hyperbolic orbit to a barycentric distance of 1,000 au. This approximation (that the solar system is a point mass with all of its mass at the barycenter) should be accurate to about one part in $10^9$, since the solar system's quadrupole term goes like $r^{-3}$. Once we have performed the analytic propagation of the rock, we use NASA's development ephemerides to initialize the solar system at a random date in a 200--year range around the arbitrarily chosen Julian Date $2459010.5$. This ensures that our results are not affected by some exceptional coincidence in the initial phases of the giant planets' orbits.

When we have initialized our rock and the solar system, we use \texttt{Rebound's} \texttt{IAS15} integrator \citep{rebound} to evolve the system numerically. For each simulation, we conserve the system's total energy to better than one part in $10^{14}$---much smaller than the fraction of the system's energy attributable to the rock. Therefore we are confident in the accuracy of our integrations.

For each integration, there are three possible outcomes: the rock may be captured; undergo a collision with another body; or be ejected from the system. If at any point during the simulation the rock's energy drops below zero, we consider it to be captured and end the simulation. If the rock undergoes a collision or if the rock is unbound and exiting the solar system with a barycentric distance greater than 40 au, we end the simulation and determine that the rock was not captured. We then follow up on our captured objects, integrating for $51\%$ of an orbital period to ensure that each object is truly bound (as opposed to having a transient bound osculating semi-major axis due to the phases of the giant planets). If during our followup the object's apocenter distance exceeds 1 parsec, we consider it to be lost to cluster or galactic tides.

Current models of solar system formation predict that the giant planets formed in a more compact arrangement, and then migrated to their current orbits. To account for this we ran a set of simulations with the compact solar system model presented in \citet{tsiganis2005}. The cross section we calculate with this model differs from that calculated using the solar system at the current epoch by less than 1 percent, so our calculations should be equally applicable to the pre-and-post-instability architectures of the solar system.

\subsection{Capture Cross Section}

\begin{figure}[ht]
    \centering
    \includegraphics[width=1\textwidth]{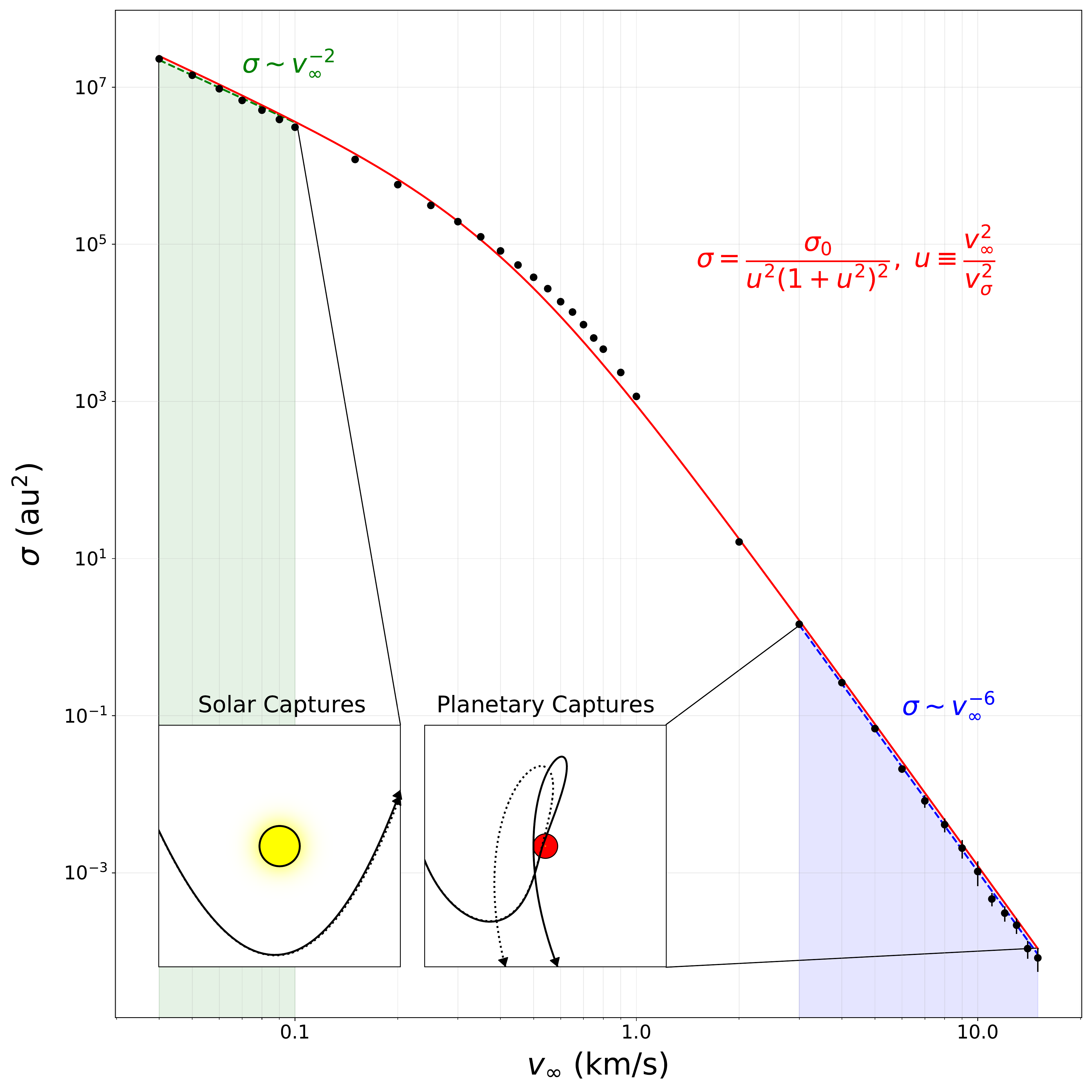}
    \caption{Capture cross section (in au$^2$) as a function of the asymptotic speed $v_{\infty}$ (in km/s). The black points represent the numerically calculated cross sections, and the corresponding error bars represent one standard deviation. The red curve represents the best fit of equation (\ref{eq:analytic_sigma}) to the data. The blue dashed line shows a power-law of the form $\sigma\sim v_{\infty}^{-6}$, as expected in the limit of high velocity. The green dashed line shows a power-law of the form $\sigma\sim v_{\infty}^{-2}$, as expected in the low velocity limit. The shaded regions indicate the parameter space where captures due the Sun (green) and planets (blue) dominate, although the boundaries are not sharp.}
    \label{fig:CrossSection}
\end{figure}

Since we sampled events uniformly in impact parameter, we can calculate the capture cross section as
\begin{equation}
    \sigma = \frac{2\pi b_{\text{max}}}{N} \sum_i b_i \delta_i
\end{equation}
where $b_{\text{max}}$ is the maximum impact parameter sampled, $N$ is the number of events, and $\delta_i$ is a Kronecker delta that is 1 if the event resulted in capture, and 0 otherwise. We display our results in Figure \ref{fig:CrossSection}. As we expect, $\sigma (v_\infty)$ goes like $v_{\infty}^{-2}$ in the low-speed limit, and like $v_{\infty}^{-6}$ in the high-speed limit. To facilitate the use of the cross section in analytic calculations we fit $\sigma(v_\infty)$ with the simple function 
\begin{equation}
    \sigma(v_\infty) = \frac{\sigma_0}{u^2 (u^2 + 1)^2}
    \label{eq:analytic_sigma}
\end{equation}
where $u \equiv v_{\infty}/v_\sigma$ and $v_\sigma$ is a velocity scale determined by the properties of the planet ejecting the rock. We find the data are best fit by parameter values $\sigma_0 = 232,250$ au$^2$ and $v_\sigma = 0.4179$ km/s. Keep in mind that these cross sections apply for capture into any bound orbit.

The scale $\sigma_0$ for the cross section obtained from fitting our numerical results can be compared to the analytic estimates of the previous section. If we evaluate equation (\ref{sunsigma}) in the high speed limit, then agreement between the analytic and numerical estimates implies that $\sigma_0$ = $64\pi\alpha^2a_J^2(m/M)^2(v_\sigma/v_J)^6$, where $v_J$ is the orbital speed of Jupiter. The expressions are equal if the dimensionless parameter $\alpha\approx0.21$. The analytic and numerical results are in agreement for all incoming speeds if we identify the scales $v_x$ and $v_\sigma$, which is equivalent to considering captures with a maximum (post-encounter) semimajor axis $a_{\rm max}\approx5090$ au. Notice also that $v_\sigma\sim{v_x}\sim(Gm/a_J)^{1/2}$ (see also Appendix \ref{sec:dimension}).  Similarly, equation (\ref{plansigma}) agrees with the numerical result in the high speed limit if the dimensionless parameter $\gamma\approx5.8$.

Note that the ensemble of numerical simulations is confined to speeds $v_\infty \le 15$ km/s. This upper limit is invoked for a number of reasons. Due to the steep power-law fall-off of the capture cross section, relatively few capture events take place at higher speeds, so additional computation leads to diminishing returns. In addition to the steep dependence with $v_\infty$, the numerical data indicate that the power-law begins to break at a comparable speed. Some type of break is expected: For $v_\infty$ greater than $\sim10$ km/s, capture by close encounters with the Sun becomes ineffective (see Appendix \ref{sec:bound}). It is noteworthy that the capture cross section at $v_\infty\approx13$ km/s is comparable to the geometrical area of the Sun ($\sim7\times10^{-5}$ au$^2$). For larger encounter 
speeds, incoming rocky bodies are thus more likely to collide with the Sun than be captured into a bound orbit.\footnote{For completeness, we note that due to gravitational focusing, the collision cross section with the Sun is larger than the capture cross section for speeds $v_\infty>2-3$ km/s.} 

\section{Analysis of Captured Objects}
\label{sec:analysis} 

\begin{figure}[ht]
    \centering
    \includegraphics[width=1\textwidth]{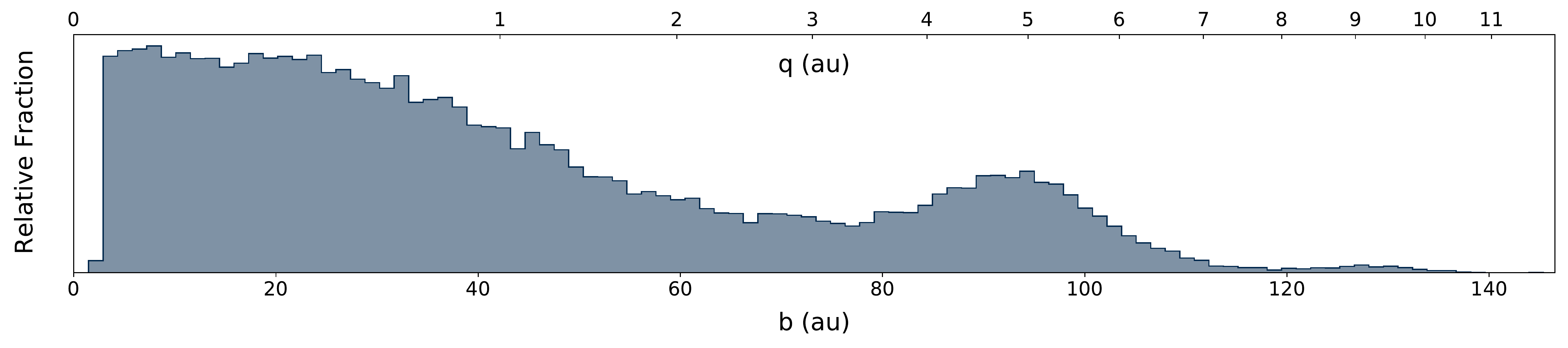}
    \includegraphics[width=1\textwidth]{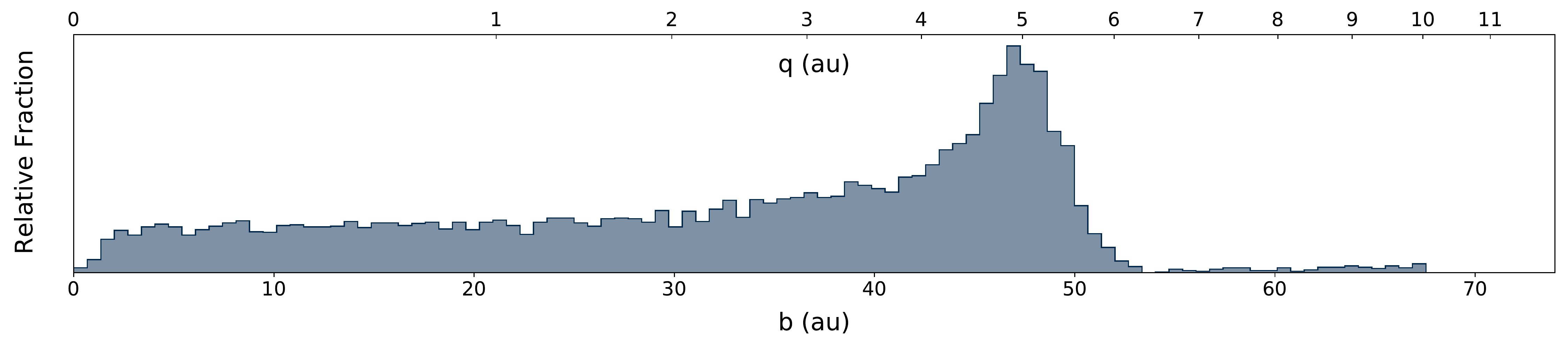}
    \caption{Histograms of the (unperturbed) impact parameter distribution of captured objects for asymptotic speeds $v_\infty$ of 1 km/s (top) and 2 km/s (bottom). For convenience, we also indicate the pericenter distance of the unperturbed orbit. Both plots show relative peaks at pericenter distances corresponding to the orbits of Jupiter and Saturn. As $v_{\infty}$ increases, close encounters with the giant planets become more important for capture.}
    \label{fig:b}
\end{figure}

In this section we examine the orbital elements of our captured objects to gain insight into the mechanics of the capture process. In Figure \ref{fig:b}, we show the impact parameter (and pericenter distance) distribution of the unperturbed orbits of our captured objects for asymptotic speeds $v_{\infty}$ of 1 and 2 km/s. Each histogram displays a clear relative peak at the pericenter distances corresponding to the orbit of Jupiter, along with a much smaller peak for the orbit of Saturn. Comparison of the two histograms indicates that somewhere between 1 and 2 km/s, the dominant capture process switches from that due to the motion of the solar system barycenter to close encounters with a giant planet (especially Jupiter).

\begin{figure}[ht]
    \centering
    \includegraphics[width=1\textwidth]{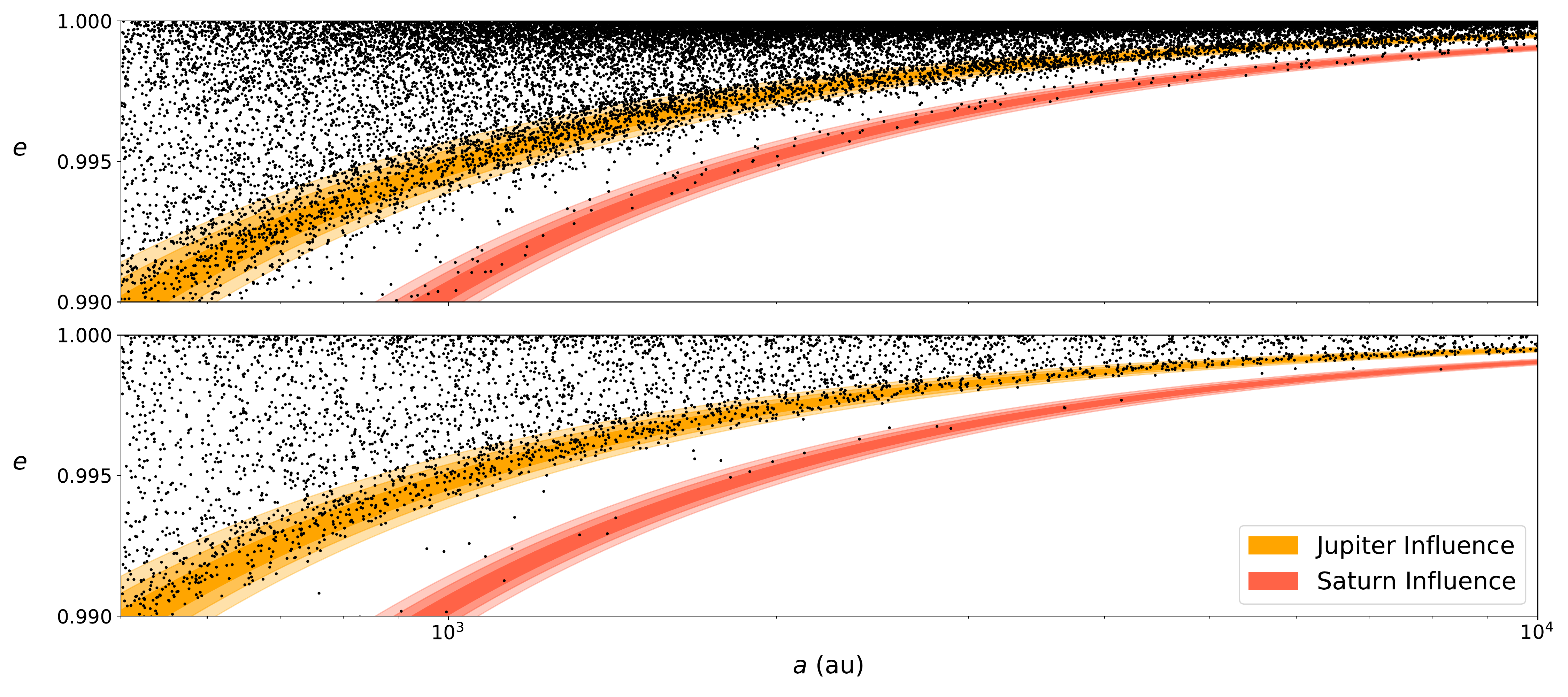}
    \caption{Post-capture eccentricity versus semi-major axis of captured objects for incoming speeds of 1 km/s (top) and at 2 km/s (bottom). The orange and red regions correspond to integer multiples of the radius of influence centered at the the semimajor axis of the orbits of Jupiter and Saturn,  respectively.}
    \label{fig:aeq}
\end{figure}

In Figure \ref{fig:aeq}, we show the post-capture eccentricity as a function of semi-major axis for a subset of the captured objects with $v_{\infty}$ of 1 km/s (top panel) and 2 km/s (bottom panel). The figure also includes equi-pericenter curves corresponding to integer multiples of the spheres of influence of Jupiter and Saturn. The numerical results for captures display a relative overdensity of points with pericenter distances at Jupiter and Saturn, indicating that these captures are (likely) attributable to close encounters.

\begin{figure}[ht]
    \centering
    \includegraphics[width=1\textwidth]{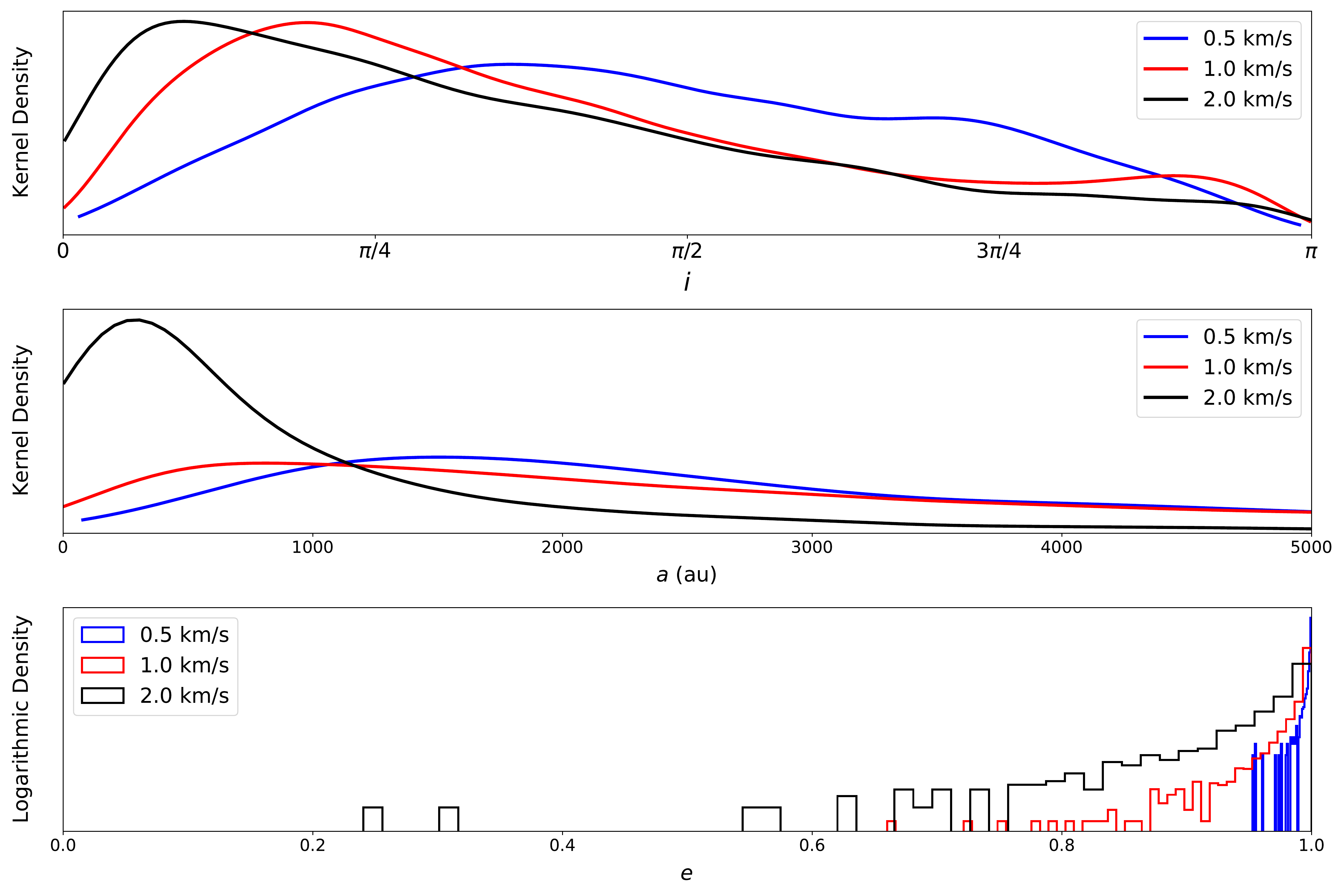}
    \caption{(Top) Gaussian kernel density estimate of the post-capture inclination distribution (with $i$ measured from the ecliptic plane) of captured objects at speeds $v_\infty$ = 0.5, 1, and 2 km/s. (Center) Gaussian kernel density estimate of the post-capture semi-major axis distribution of captured objects for $v_\infty$ = 0.5, 1, and 2 km/s. (Bottom) Relative fraction (in logarithmic scale) of the post-capture eccentricity distribution of captured objects for $v_\infty$ = 0.5, 1, and 2 km/s. Note that these curves represent probability distribution functions; there will be fewer captures in the high-speed case than in the low-speed case.}
    \label{fig:ainc}
\end{figure}

In Figure \ref{fig:ainc}, we show the kernel density representations for the post-capture inclination semi-major axis, and eccentricity for $v_{\infty}$ = 0.5, 1, and 2 km/s. While captures become increasingly rare for higher-velocity events, the resulting semi-major axes of the captured objects are typically smaller than those for objects captured in low-velocity events. This trend is important for assessing object retention, as captured bodies with semi-major axes $a \gtrsim 1000$ au are more likely to be stripped from the solar system by interacting with passing stars (in the solar birth cluster) or the galactic tides (in the field). 

It is noteworthy that capture events readily produce highly-inclined and even retrograde objects. This finding indicates that capture is yet another potential channel for the production of the observed populations of highly-inclined and retrograde centaurs, which are currently best explained by the putative Planet Nine \citep{pninereview}. However, it is important to note that the orbits of the captured objects will evolve over time. As a result, the captured objects do not represent a long-term stable population. As the orbits of the captured objects evolve, some will become more eccentric until they collide with the Sun; some will undergo scattering events or interactions with the Galactic tides and be ejected from the solar system; some will be frozen into the inner Oort cloud by passing stars; and some will continue to evolve on stable or quasi-stable orbits (either by becoming caught in resonances with the giant planets, or by achieving orbits that otherwise avoid close encounters). 

Because most capture events resulted in high-eccentricity orbits, we have rather low statistics for small values of eccentricity. Despite the data limitations, though, it is clear that as $v_{\infty}$ increases, the low-eccentricity tail of the distribution becomes fatter.

\section{Applications} 
\label{sec:apply}  

\subsection{Velocity Averaged Cross Sections} 
\label{sec:meansigma} 

This paper determines the velocity dependent cross section $\sigma(v_\infty)$, which can be fit with a function of the form  
\be
\sigma(v_\infty) = {\sigma_0 \over u^2 (1 + u^2)^2} \qquad 
{\rm where} \qquad u \equiv {v_\infty \over v_\sigma} \,.
\label{sigofv} 
\ee
The capture rate for rocky bodies by our solar system is given by 
\be
\Gamma = n_R \langle \sigma v_\infty \rangle \,, 
\label{rockrate} 
\ee
where $n_R$ is the number density of rocks that the solar system encounters. The capture rate depends on the velocity-averaged cross section, which is given by the integral 
\be
\langle \sigma v_\infty \rangle = \int_0^\infty v_\infty f(v_\infty) \sigma(v_\infty) dv_\infty \,,
\ee
where $f(v_\infty)$ is the distribution of encounter velocities of the rocky bodies.

The distribution of relative speeds $f(v_\infty)$ depends on the environment. In the solar birth cluster, $f(v_\infty)$ is determined by the processes that eject the rocky bodies from their original planetary systems. In general, the clusters are not sufficiently long-lived for the rocks to attain a thermal distribution of speeds. Instead, they are expected to retain the velocity distribution resulting from the ejection process. If ejection occurs through scattering interactions with giant planets, then $f(v_\infty)$ takes the approximate form 
\be
f(v_\infty) = {4 v_\infty/v_p \over (1 + v_\infty^2/v_p^2)^3} \,, 
\label{veldist} 
\ee
where the velocity scale $v_p^2\approx GM_\ast/a_p$, where $a_p$ is the semi-major axis of the planet that scatters the rocks (e.g., see \citealt{moorhead} for a derivation).

Note that the distribution (\ref{veldist}) is normalized over the entire interval $0\le{v_\infty}\le\infty$. In practice, the distribution will have a maximum value determined by the escape speed from the planets that scatter the rocky bodies. Notice also that the full  distribution will be a convolution of the distribution of ejection speeds from each planet that scatters rocky material. As an approximation, we consider only a single distribution and interpret the velocity scale $v_p$ as a typical value. As a result, $v_p$ is expected to be comparable to the orbit speed of outer planets, i.e., $v_p \sim 10$ km/s. Finally, we are assuming that equation (\ref{veldist}) corresponds to the distribution of {\it relative} speeds between the rocks and the solar system (e.g., see the discussion of \citealt{binneytremaine}). 

Putting the above considerations together, we can write the velocity averaged cross section in the form 
\be
\langle \sigma v_\infty \rangle = \sigma_0 {v_\sigma^3 \over v_p^2} 
\int_0^\infty  {4 du \over (1+\eta^2 u^2)^3 (1+u^2)^2} =
\sigma_0 {v_\sigma^3 \over v_p^2} I(\eta) \,,
\label{sigvelx} 
\ee
where $u=v_\infty/v_\sigma$ (as before), we have defined 
$\eta\equiv v_\sigma/v_p$, and where the second equality defines the integral function $I(\eta)$. The dimensionless function $I(\eta)$ can be evaluated to obtain 
\be
I(\eta) = \pi {(1 + \eta)(1 + 3\eta) + 3\eta^3/4 \over (1+\eta)^4} \,.
\ee
Note that $I\to\pi$ in the limit $\eta\to0$, and in practice $\eta\sim1/10$. As a result, a good approximation for the capture cross section takes the form 
\be
\langle \sigma v_\infty \rangle \approx 
\pi \sigma_0 {v_\sigma^3 \over v_p^2} \,. 
\ee  

For comparison, we can determine the velocity averaged cross section for the scenario where the cluster rocks are virialized and have the same (Maxwellian) velocity distribution as the stars.\footnote{In general, we expect the stars to reach virial equilibrium much faster than the rocky ejecta. The stars start their cluster trajectories with sub-virial speeds, but then fall toward the cluster core where interactions take place, and equilibrium is rapidly realized (in a few Myr, e.g., \citealt{proszkow}). In contrast, the rocks are ejected with speeds much larger than the virial speed and have little chance for interactions to slow them down. Moreover, the stellar virialization starts as soon as stars form, whereas the planet formation and the subsequent ejection of rocks occurs many Myr later.} In this limit, $\langle\sigma{v_\infty}\rangle$ can be written in the form  
\be
\langle \sigma v_\infty \rangle = \sigma_0 {v_\sigma^2 \over s} 
\sqrt{2\over\pi} \int_0^\infty {v_\infty dv_\infty \over s^2} 
{\exp[-v_\infty^2/2s^2] \over (1 + v_\infty^2/v_\sigma^2)^2} \,, 
\ee
where $s$ is the velocity dispersion of the distribution. Note that the value of $s$ for the distribution of relative speeds is larger than the value $s_0$ for the velocity distribution of the stars in the clusters ($s=\sqrt{2}s_0$).  Here we define the variable $w\equiv v_\infty/s$ and the parameter $\xi=s/v_\sigma$, so that 
\be
\langle \sigma v_\infty \rangle = \sigma_0 {v_\sigma^2 \over s} 
\sqrt{2\over\pi} \int_0^\infty wdw 
{\exp[-w^2/2] \over (1 + \xi^2 w^2)^2} \equiv 
\sigma_0 {v_\sigma^2 \over s} \sqrt{2\over\pi} J(\xi) \,, 
\label{sigvelg} 
\ee
where the second equality defines the integral function $J(\xi)$.  The exact form for $J(\xi)$ can be found. If we define
$\mu=1/(2\xi^2)=v_\sigma^2/(2s^2)$, then 
\be
J (\xi) = J(\mu) = 
\mu \left[ 1 - \mu\,{\rm e}^\mu\,E_1(\mu) \right]\,,
\label{jexact} 
\ee
where $E_1(x)$ is the exponential integral \citep{abrasteg}. In the limit $\xi\to0$ ($\mu\to\infty$), the function $J=1$; in the opposite limit $\xi\gg1$ ($\mu\to0$), $J=\mu=1/(2\xi^2)$. A good working approximation for the 
cross section of equation (\ref{sigvelg}) thus has the form 
\be
\langle \sigma v_\infty \rangle \approx \sigma_0 {v_\sigma^4 \over s} 
{\sqrt{2/\pi} \over v_\sigma^2 + 2 s^2}  \,,
\ee
which is exact in the limits and has a relative error less than $\sim20\%$ over the entire range $0\le\xi\le\infty$. One can use equations (\ref{sigvelg}) and (\ref{jexact}) if higher accuracy is required. 

\subsection{Rock Capture in the Birth Cluster} 
\label{sec:rockmass} 

Using the results derived above, we can estimate the total mass in rocky bodies that were captured while the Sun remained in its birth cluster. The capture rate is given by equation (\ref{rockrate}), with velocity-averaged cross section specified through equations (\ref{sigvelg}) and (\ref{jexact}).  The rocky bodies will have a distribution of sizes $g(R)$, which is defined here such that 
\be
n_R = \int_0^\infty g(R) dR \qquad {\rm and} \qquad 
\rho_R = \int_0^\infty g(R) m(R) dR \,,
\ee
where $m(R)$ is the mass of the rock as a function of its size. With these definitions, the capture rate $\Gamma$ can be converted into a mass accretion rate given by 
\be
{\dot M} = \rho_R \langle \sigma v_\infty \rangle \,, 
\ee
where $\rho_R$ is the mass density of the cluster in the form of rocks. Given that each planetary system in the cluster is expected to eject a few Earth masses of rocky material (e.g., \citealt{RiceLaughlin2019}), the density $\rho_R$ is given by  
\be
\rho_R = {\alpha M_\oplus N_\ast \over V} 
= \alpha M_\oplus n_\ast \,,
\ee
where $\alpha$ is a dimensionless factor of order unity and $n_\ast$ is the number density of stars. For completeness, note that the inclusion of icy planetesimals will increase this density estimate. In any case, the total mass in rocky bodies captured by the solar system during its cluster phase can be written in the form  
\be
(\Delta M)_R = \alpha M_\oplus 
\left[ \int_0^\infty n_\ast dt \right] 
\langle \sigma v_\infty \rangle \equiv
\alpha M_\oplus \langle n_\ast \rangle \tau 
\langle \sigma v_\infty \rangle \,. 
\label{mtotal} 
\ee
The final equality defines the mean density of the cluster, where $\tau$ is its effective lifetime. A number of studies have found upper bounds on the product $\langle{n_\ast}\rangle\tau$ by requiring that the solar system is not overly disrupted, including considerations of the planetary orbits \citep{al2001,adams2010,gdawg2015}, the Kuiper Belt \citep{newdawg}, and the orientation of the plane of the cold classicals \citep{yuri}. This work indicates that the product is bounded by  $\langle{n_\ast}\rangle\tau\lta$ $2\times10^4$ pc$^{-3}$ Myr. 

If we take $v_\sigma$ = 0.5 km/s, $v_p$ = 10 km/s, and $\sigma_0$ = $2\times10^5$ au$^2$, then the velocity averaged capture cross section becomes $\langle\sigma{v_\infty}\rangle$ $\approx$ 800 au$^2$ km/s. With this cross section, the total mass in captured rocky material from equation (\ref{mtotal}) is about $(\Delta M)_R\sim10^{-3}M_\oplus$. Of course, most of this material will be ejected back into the cluster or the field. The retention rate of material in the inner Oort cloud is $\sim 1\%$ \citep{Brasser2006}, so we would expect $\sim10^{-5}M_\oplus$ to be captured in the inner Oort cloud.

Note that if rocks ejected from planetary systems in clusters follow the velocity distribution of equation (\ref{veldist}), then some fraction of the material will leave the cluster during its first crossing. The high speed tail of the velocity distribution will thus be de-populated. In practice, however, most of the capture events arise from the low-speed portion of the distribution, so that the correction for the loss of high speed material is modest. 

We can also estimate the mass of rocks captured while the solar system is in the field. In this case, we expect the rocky material to encounter the solar system with a velocity distribution comparable to that of the field stars, i.e., a Gaussian distribution with $s\sim40$ km/s. In this case, the velocity averaged cross section $\langle\sigma{v_\infty}\rangle\approx0.08$ au$^2$ km/s. If we also assume that each planetary system ejected the same mass in rocks during its formative phases, then the density of rocky material will be proportional to the stellar density (we are thus assuming negligible losses). As a result, the product $\langle{n_\ast}\rangle\tau\sim460$ pc$^{-3}$ Myr, and the expected mass in captured rocks is about $(\Delta M)_R\approx$ $2\times10^{-9}M_\oplus$. Using the approximate retention rate of $1\%$, we would expect only $\sim 2\times10^{-11}M_\oplus$ of these rocks to remain in the inner Oort cloud. This inventory of captured alien material from the field is exceedingly small, roughly the equivalent of one 5 km body. Rock capture during the birth cluster phase is thus expected to produce the dominant contribution (by roughly a factor of one million). These latter objects are expected to have radiogenic ages comparable to ordinary solar system bodies, but might be identified by different (unusual) chemical composition.

Note that the values presented here are highly approximate. Not all of the rocks will be captured in the inner Oort cloud, so that the retention fraction could be smaller than assumed here (most of the captured interstellar bodies initially have Jupiter-crossing orbits, whereas the planetesimals in the Oort cloud could have different origins). In any case, most of the captured objects will be ejected, and some will eventually collide with the Sun. Although these calculations provide working order-of-magnitude estimates, in forthcoming work we will refine these projections by numerically investigating the long-term behavior of the captured bodies from this work. 

\section{Conclusions} 
\label{sec:conclude} 

This paper has revisited the problem of capturing interstellar objects on initially hyperbolic trajectories into bound states. Using an ensemble of 500 million numerical fly-by simulations, the main result of this study is the determination of our solar system's capture cross section as a function of encounter  speed (see Figure \ref{fig:CrossSection}). The resulting capture cross section shows the power-law velocity dependence $\sigma \sim v_\infty^{-2}$ in the limit of low speeds and the dependence $\sigma \sim v_\infty^{-6}$ in the limit of high speeds. The capture cross section $\sigma(v_\infty)$ over the entire range of asymptotic speeds can be fit with the function given in equation (\ref{sigofv}). 

This paper also presents an analytic treatment of the capture problem using the approximation of matched conics and the (inverse) gravitational slingshot effect (Section \ref{sec:capture}). These arguments show that capture by both close encounters with the Sun (Section \ref{sec:starslingshot}), and by close encounters with a giant planet (Section \ref{sec:planslingshot}), have the same nearly velocity dependence as that seen in the numerical simulations (namely $\sigma\sim v_\infty^{-2}$ at low speeds and $\sigma\sim v_\infty^{-6}$ at high speeds). 

The capture events can be classified as either close encounters with the Sun or close encounters with giant planets. At low speeds, encounters with the Sun dominate the capture cross section. At higher speeds, close encounters with Jupiter dominate. Close encounters with the other giant planets contribute to the cross section, but do not dominate the dynamics. More specifically, for the particular case of $v_\infty$ = 1 km/s, capture events due to close encounters with Jupiter are $\sim100$ times more likely than captures due to Saturn. The frequency of close encounters with Uranus and Neptune are smaller (than for Saturn) by an additional factor of $\sim100$. 

With the capture cross section as a function of velocity  specified, the effective mean cross section $\langle v \sigma \rangle$/$\langle{v}\rangle$ can be determined for any distribution of encounter speeds. For the case of a Maxwellian distribution and a power-law distribution motivated by rock ejection, the mean cross section can be evaluated analytically (see Section \ref{sec:meansigma}). 

Finally, as an application of the capture cross section, we estimate the total mass $(\Delta M)_R$ in the Oort cloud that originates from other planetary solar systems (Section \ref{sec:rockmass}). The mass accreted while the Sun lived within its birth cluster is of order $(\Delta M)_R\sim10^{-5}M_\oplus$, about a million time larger than the mass subsequently accreted from the field. 

Although the capture cross section for the solar system is now well-characterized, many avenues for future research remain. The simulations of this paper consider the capture of interstellar objects and the resulting cross sections include all capture events, independent of their residence time in the solar system as bound objects. Future work should determine how long captured bodies can remain bound to the Sun, since many such objects are expected to be ejected from the system or to collide with other solar system members. The residence time (ejection time) should thus be determined for each type of orbit displayed by the captured objects. With these results in place, one can make a refined estimate of the current population of alien objects in the solar system, along with their expected orbital properties. 

\acknowledgements

We would like to thank David Gerdes, Hsing-Wen Lin, and Larissa Markwardt for helpful discussions during the preparation of this manuscript. 

This material is based upon work supported by the National Aeronautics and Space Administration under Grant No. NNX17AF21G issued through the SSO Planetary Astronomy Program and by the National Science Foundation under Grant No. AST-2009096.

\appendix
\section{Upper Bound on Incoming Speed for Capture} 
\label{sec:bound} 

In this Appendix, we find upper limits on the asymptotic speed, i.e., the largest speed for which a rock can be captured by the solar system. We consider both close encounters with the Sun and with one of the giant planets.

We start with the capture criterion of equation (\ref{vlimit}), 
which we reproduce here,
\be
v_\infty^2 < 4Uv \cos\theta - 4U^2 < 4Uv \,, 
\ee
where the second inequality follows because $\cos\theta<1$  and $4U^2>0$. The largest speed $U$ possible for the Sun, relative to the center of mass, is the orbit speed due to the giant planets, where 
\be
U = {m \over M} \left({G(M+m)\over a_p}\right)^{1/2} 
\approx {m \over M} \left({GM\over a_p}\right)^{1/2} \,,
\ee
where the second expression is consistent with the ordering approximation and where we have included the reflex speed due to only one planet. The maximum possible speed is given by the sum of the contributions of all of the planets. We can account for this complication by writing the limit in the form 
\be
U < {2m_J \over M} \left({GM\over a_J}\right)^{1/2} \,. 
\ee
Thus far, our upper bound has the form 
\be
v_\infty^2 < 8 {m_J \over M} \left({GM\over a_J}\right)^{1/2} v 
= 8 {m_J \over M} \left({GM\over a_J}\right)^{1/2} 
\left[ v_\infty^2 + {2GM \over R_\odot} \right]^{1/2} \,,
\ee
where we have taken the minimum distance (and maximum speed) of the orbit to be given by the radius of the Sun. If we drop the first term in square brackets, the expression simplifies to the form 
\be
v_\infty^2 < 8 {m_J \over M} \left({GM\over a_J}\right)^{1/2} 
\left( {2GM \over R_\odot} \right)^{1/2} = 8\sqrt{2} {G m_J \over (a_J R_\odot)^{1/2}} 
\approx (8\,{\rm km/s})^2 \,.
\label{vbound} 
\ee
To derive a rigorous version of the upper bound, we define two velocity components 
\be
u \equiv 8 {m_J \over M} \left({GM\over a_J}\right)^{1/2} 
\qquad {\rm and} \qquad 
w \equiv \left({2GM \over R_\odot}\right)^{1/2} \,, 
\ee
so that the limit takes the form 
\be
v_\infty^2 < uw \left[1 + {u^2 \over 4w^2} \right]^{1/2} + 
{u^2 \over 2} \,. 
\label{full} 
\ee
Since $u$ is {\it of order} the orbital speed of the Sun ({specifically, a small fraction of 1 km/s}) and $w$ is the escape speed of the Sun ($\sim620$ km/s), we find $u\ll{w}$. In this limit, the full expression of equation (\ref{full}) reduces to the form $v_\infty^2 < uw$, which corresponds to the approximation of equation  (\ref{vbound}). 

We can also consider the case where the incoming rock enters into the sphere of influence of a planet and loses energy through an inverse gravitational assist from the encounter.  In order for the rock to be captured by the solar system, the final speed must be sufficiently small, i.e., 
\be
v_2^2 < 2\vorb^2 \,, 
\ee
where the orbital speed $\vorb$ of the planet is a measure of the depth of the gravitational potential well at the location of the planet (the location of the encounter). We also assume that the incoming rock obeys conservation of energy so that 
\be
{1 \over 2} v_\infty^2 = E = {1\over2} v^2 - {GM\over r} 
\approx {1\over2} v^2 - \vorb^2 \qquad \Rightarrow \qquad 
v^2 = v_\infty^2 + 2\vorb^2 \,. 
\ee
Using this expression in the result for the post-encounter speed from before, we obtain 
\be
v_\infty^2 + 4U^2 < 4U \cos\theta [v_\infty^2 + 2\vorb^2]^{1/2} \,.
\ee
If we take $U=\vorb\cos\phi$, this expression can be rewritten in the form 
\be
v_\infty^4 < 8 \vorb^2 v_\infty^2 \cos^2\phi
(2\cos^2\theta - 1) + 16 \vorb^4 \cos^2\phi
(2\cos^2\theta - \cos^2\phi)\,.  
\ee
We can thus obtain an upper limit on the asymptotic speed for an object to be captured by taking $\cos\theta=1=\cos\phi$, i.e.,
\be
v_\infty^4 < 8 \vorb^2 v_\infty^2 + 16 \vorb^4 \,, 
\ee
which leads to the bound 
\be
v_\infty < 2\vorb (1 + \sqrt{2})^{1/2} \,.  
\ee
The fastest orbit is that of Jupiter, where $\vorb\approx$ 13 km/s, so we have the limit $v_\infty\lta40$ km/s.

We thus find that the maximum possible speed (for capture) is larger for the channel involving close approaches to planets (in particular Jupiter) than for close approaches to the Sun.  On the other hand, planet encounters are expected to occur much less frequently. Note that these limits do not account for effects such as radiation pressure or atmospheric drag. It is therefore possible in principle to capture rocks with larger $v_\infty$, but such events would be exceptionally rare.

\section{Rock Capture by Circumstellar Disks} 
\label{sec:gascap} 

Rock capture could also take place due to gas drag if interstellar bodies enter the solar nebula while it retains its gaseous component. This Appendix explores the efficacy of this process. 

If a rock has speed $v_0$ when it enters the gaseous region of the disk, it will have final speed $v_f$ given by
\be 
v_f = v_0 \exp [ - \rho A \ell / m ] \,,
\ee
where $\rho$ is the gas density, $A$ is the rock area, $m$ is the rock mass, and $\ell$ is the length traveled.  If a rock comes from infinity with speed $v_\infty$, then it will have a larger velocity when it hits the disk, so that $v_0$ is given by
\be
v_0^2 = v_\infty^2 + {2GM \over r } \,. 
\ee
In order for the rock to lose enough energy to enter into a 
bound orbit, the final speed must be less than the limit
\be
v_f^2 = \left[v_\infty^2 + {2GM \over r } \right] 
\exp\left[ - {2\rho A\ell \over m} \right] < {2GM \over r} \,. 
\ee
The capture criterion thus becomes 
\be
{2 \rho A \ell \over m} > \ln \left[ 1 + 
{v_\infty^2 r \over 2GM} \right] \,. 
\ee
If the rock passes through the disk vertically, then $\rho\ell=\Sigma$, where the surface density for the 
solar nebula can be written in the form 
\be
\Sigma(r) = \Sigma_1 
\left({1 {\rm au} \over r}\right)^{3/2} \,,
\ee
where $\Sigma_1 \approx 3000$ g/cm$^2$.  If the rock passes through the nebula at an arbitrary angle $\theta$, then $\rho\ell=\Sigma/\cos\theta$. We can also write 
$A = \pi R^2$ and $m = \rho_R (4\pi/3) R^3$ where $R$ is 
the size of the rock. If we work in the low speed limit with $v_\infty$ = 1 km/s, and write $R$ in units km and $r$ in units of au, then the criterion for capture reduces to the simple form 
\be
{36 \over \cos\theta} > R r^{5/2} \,.
\ee
For the fiducial case where $\cos\theta=1$, the solar nebula can thus capture incoming rocks with radii $R\lta1$ km at radial location $r=1$ au.  Somewhat larger rocks can be captured for typical inclined trajectories (e.g., the limit becomes $R\lta2$ km for $\cos\theta=1/2$). Smaller rocks are readily captured, and rocks as large as dwarf planets are highly unlikely to be captured (see also  \citealt{brasser2007}). For incoming trajectories that are confined to the plane of the disk, the above treatment must be modified to include the disk structure (i.e., the above treatment does not apply in the limit $\cos\theta\to0$).

\section{Dimensional Analysis} 
\label{sec:dimension} 

The capture problem has more than one dimensionless field, so that the cross section of interest cannot be directly determined from dimensional analysis. Nonetheless, such an analysis is useful for understanding the result. 

Consider the solar system to consist of only the Sun-Jupiter binary. The physical variables required to characterize the systems are thus the masses $(M,m)$ and the orbital radius of Jupiter $a_J$. Note that the eccentricity of the Jovian orbit is too small to matter, and the radii of the bodies also do not contribute. The other variables in the problem are the gravitational constant $G$ and the speed of the incoming rock $v_\infty$. Here we want to determine the cross section as a function of these variables. We thus want to determine the impact parameter $b$ of the incoming orbits that allow for capture. 

The first estimate for the cross section is thus 
\be
\sigma_1 = \pi a_J^2 \,. 
\ee
The remaining (relevant) variables can be combined to construct two dimensionless fields, i.e., 
\be
\Lambda_1 \equiv {GM \over v_\infty^2 a_J} \qquad {\rm and} 
\qquad \Lambda_2 \equiv {Gm \over v_\infty^2 a_J} \,. 
\ee 
Alternatively, one could define the second field to be the mass ratio $m/M$, and the quantity $\Lambda_2=\Lambda_1(m/M)$. At low speeds, the capture cross section has the form 
\be
\sigma_2 = \pi a_J^2 \Lambda_1 \,, 
\ee
whereas at high speeds the cross section becomes
\be
\sigma_3 = \pi a_J^2 \Lambda_1 \Lambda_2^2 \,. 
\ee
We can combine all of the above results to write the capture cross section in terms of the dimensionless fields of the problem: 
\be
\sigma_4 = \pi a_J^2 
{\Lambda_1 \Lambda_2^2 \over (1 + \Lambda_2)^2 } \,. 
\ee

\bibliographystyle{aasjournal}
\bibliography{references}

\end{document}